\documentclass[
    ,final            
  ]
  {aipproc}

\layoutstyle{6x9}

\newcommand{\be}{\begin{equation}}
\newcommand{\ee}{\vspace{0cm} \end{equation}}
\newcommand\la{\lower0.6ex\vbox{\hbox{\ensuremath{\buildrel{\textstyle<}\over{\sim}\ }}}}
\newcommand\ga{\lower0.6ex\vbox{\hbox{\ensuremath{\buildrel{\textstyle>}\over{\sim}\ }}}}
\newcommand{\obh}{\ensuremath{\Omega_{\rm B} h^2\;}}

\newcommand{\omb}{\ensuremath{\Omega_{\rm B}\;}}

\def\eg{{\it e.g.},~}

\def\4he{$^4$He}
\def\3he{$^3$He}
\def\7li{$^7$Li}

\def\ydp{$y_{\rm DP}$~}

\def\hi{H\thinspace{$\scriptstyle{\rm I}$}}

\begin{document}

\title{TRACKING THE POST-BBN EVOLUTION OF DEUTERIUM}


\author{Gary Steigman}{
  address={Departments of Physics and Astronomy\\
  Center for Cosmology and Astro-Particle Physics\\
  The Ohio State University\\
  191 West Woodruff Avenue\\
  Columbus, OH 43210, USA}
}

\begin{abstract}
The primordial abundance of deuterium produced during Big Bang Nucleosynthesis 
(BBN) depends sensitively on the universal ratio of baryons to photons, an 
important cosmological parameter probed independently by the Cosmic Microwave 
Background (CMB) radiation.  Observations of deuterium in high-redshift, 
low-metallicity QSO Absorption Line Systems (QSOALS) provide a key baryometer, 
determining the baryon abundance at the time of BBN to a precision of $\sim 5$\%.  
Alternatively, if the CMB-determined baryon to photon ratio is used in the BBN 
calculation of the primordial abundances, the BBN-predicted deuterium abundance 
may be compared with the primordial value inferred from the QSOALS, testing the 
standard cosmological model.  In the post-BBN universe, as gas is cycled through 
stars, deuterium is only destroyed so that its abundance measured anytime, 
anywhere in the Universe, bounds the primordial abundance from below.  Constraints
on models of post-BBN Galactic chemical evolution follow from a comparison of
the relic deuterium abundance with the FUSE-inferred deuterium abundances in 
the chemically enriched, stellar processed material of the local ISM.
\end{abstract}

\maketitle

\section{Introduction}

Of the light nuclides synthesized in astrophysically interesting abundances
(D, \3he, \4he, \7li) during Big Bang Nucleosynthesis (BBN), the post-BBN
evolution of deuterium is the simplest.  As gas is cycled through stars,
deuterium is only destroyed~\cite{els}, so that its abundance, observed
anywhere in the Universe, at any time in its evolution, is constrained
to be bounded from above by its primordial, BBN abundance: (D/H)$_{\rm
OBS}~\leq~$(D/H)$_{\rm P}$.  In systems at high redshift and/or with very
low metallicity, the observed deuterium abundance should approach its 
primordial value.  Its simple post-BBN evolution, along with the sensitivity 
of its predicted BBN abundance to the baryon to photon ratio $\eta_{\rm B} 
\equiv n_{\rm B}/n_{\gamma}$ ((D/H)$_{\rm P} \propto \eta_{\rm B}^{-1.6}$),
identifies deuterium as the baryometer of choice.  The baryon mass density
parameter, \omb (the fraction of the present critical mass density contributed
by baryons), is related to $\eta_{\rm B}$ and the present value of the
Hubble parameter, $H_{0} \equiv 100h$~kms$^{-1}$Mpc$^{-1}$, by \obh =
$\eta_{10}/274$, where $\eta_{10} \equiv 10^{10}\eta_{\rm B}$.  Given
the dependence of (D/H)$_{\rm P}$ on $\eta_{\rm B}$, observations which
constrain the relic deuterium abundance to, say, $\sim 10$\%, lead to a
$\sim 6$\% measurement of the baryon to photon ratio when the Universe
was only a few minutes old.  Alternatively, the baryon to photon ratio
determined by observations of the cosmic microwave background (CMB), which
probe a time when the Universe was some 400 thousand years old, may be 
used in the BBN calculation to predict the primordial D abundance, which 
may then be compared to the relic value inferred from observations of 
deuterium at high redshifts in systems of very low metallicity, testing
the standard model of cosmology.

In the post-BBN Universe, the D ``astration'' factor, $f_{\rm D}~\equiv
$~(D/H)$_{\rm P}$/(D/H), measures the virgin fraction ($1/f_{\rm D}$);
$1/f_{\rm D}$ is the fraction of gas which has never been processed
through stars.  Due to stellar nucleosynthesis, as the metallicity, 
$Z$, in a system increases, the deuterium abundance should decrease, 
suggesting that, in the absence of dust depletion, the observed values 
of D/H and $Z$ should be anti-correlated.

\section{BBN And The Primordial Abundance Of Deuterium}

\begin{figure}
  \includegraphics[height=.35\textheight]{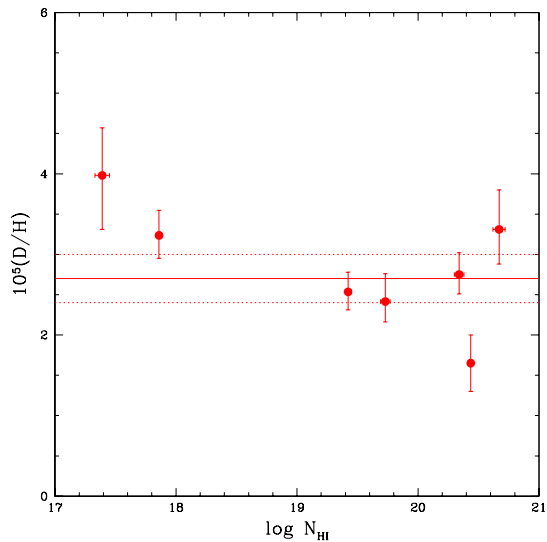}
  \caption{Deuterium abundances, the D to H ratios by number, derived
  from observations of high redshift, low metallicity QSO Absorption
  Line Systems, are shown as a function of the corresponding neutral
  hydrogen, \hi, column densities.  The solid line shows the weighted
  mean of the D/H ratios and the dashed lines indicate the $\pm 1\sigma$ 
  error estimates; see the text.}
  \label{fig:dpobs}
\end{figure}

In Figure~\ref{fig:dpobs} are shown the deuterium abundances inferred
from observations of seven, high redshift, low metallicity QSO Absorption
Line Systems (QSOALS)~\cite{pettini}.  The weighted mean of these abundance
determinations provides an estimate of the primordial deuterium abundance,
\ydp $\equiv 10^{5}$(D/H)$_{\rm P} = 2.7 \pm 0.2$.  Since the dispersion
of the individual abundance determinations around the mean value is very
large, the formal error in the mean has been multiplied by the square root
of the reduced $\chi^{2}$ to provide a more realistic error estimate.

From BBN~\cite{bbn}, for \ydp $= 2.7 \pm 0.2$, the baryon abundance is
determined, at the $\sim 5$\% level, to be $\eta_{10} = 6.0 \pm 0.3$.
This determination, some $\sim 20$ minutes after the expansion has begun,
is in excellent agreement with the value determined independently from
the CMB, some 400 kyr later, $\eta_{10} = 6.1 \pm 0.2$ (see Simha \&
Steigman~\cite{bbn} and references therein).  If the CMB-determined 
value of the baryon abundance is used to predict the primordial abundance 
of deuterium, the result is \ydp $= 2.6 \pm 0.1$, in perfect agreement, 
within the estimated errors, with the relic abundance inferred from 
the QSOALS data.

\section{Observations of D In the ISM}

\begin{figure}
  \includegraphics[height=.355\textheight]{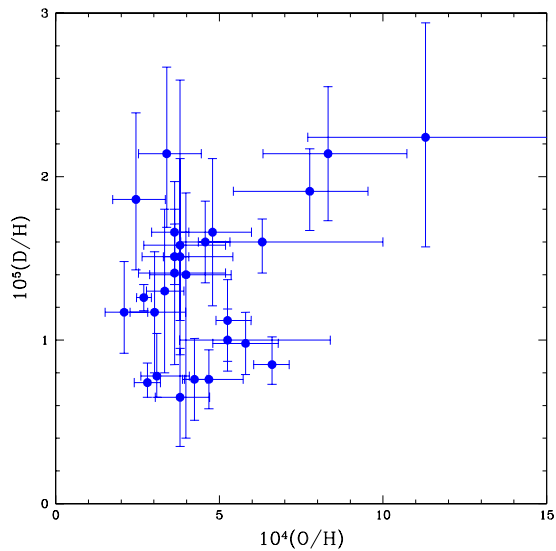}
  \caption{The FUSE-observed ISM abundances of deuterium versus those 
  of oxygen observed along the same lines of sight; from~\cite{srt}.}
  \label{fig:dvso}
\end{figure}

The conventional wisdom (ante?) has been that mixing of the ISM within 
1 -- 2 kpc of the location of the solar system is efficient so that this 
relatively local gas is well-mixed with homogenized abundances.  In particular, 
the conventional wisdom led to the expectation that observations of deuterium 
and, \eg oxygen within the local ($\sim 1$~kpc) ISM would reveal uniform 
abundances, ``the interstellar abundances'', of these (and other) elements.  
Beginning with the pioneering Copernicus observations \cite{copernicus}, 
continuing with observations provided by the IUE satellite, and the GHRS 
and STIS instruments onboard the HST, and confirmed by those of the FUSE 
mission~\cite{linsky}, the data reveal that nothing could be further from 
the truth.   As may be seen in Figure~\ref{fig:dvso}, the deuterium abundances 
inferred from observations in the relatively local ISM (see \cite{linsky} 
for details and further references) vary over a range of a factor of $\sim 
3$, while those of oxygen range over a factor of $\sim 4$.

\begin{figure}
  \includegraphics[height=.355\textheight]{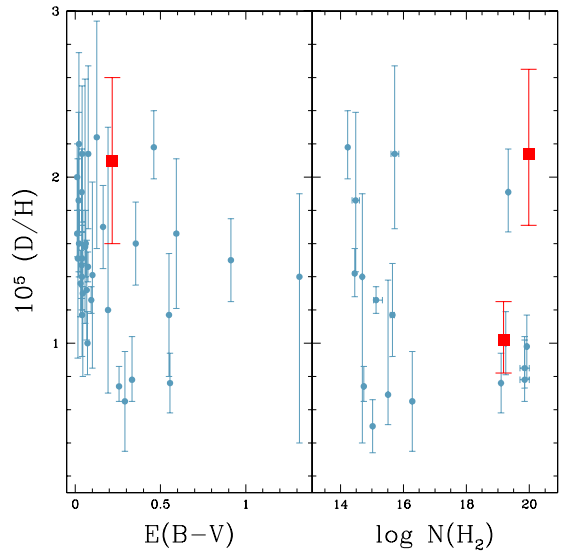}
  \caption{In the left-hand panel the FUSE observed D/H ratios are shown 
  versus the corresponding LOS values of E(B - V) and, in the right-hand 
  panel of log~N(H$_{2}$); from~\cite{srt}.}
  \label{fig:ext}
\end{figure}

\begin{figure}
  \includegraphics[height=.345\textheight]{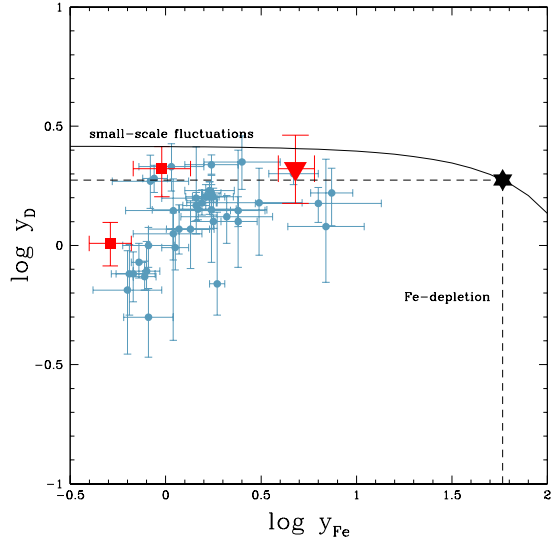}
  \caption{FUSE observed ISM D and Fe abundances ($y_{\rm D} \equiv 10^{5}$(D/H), 
  $y_{\rm Fe} \equiv 10^{6}$(Fe/H)); from~\cite{srt}.  The star shows the ISM 
  abundances predicted by the fiducial GCE model (see the text).  D and Fe 
  abundances below and to the left of the dashed lines are consistent with 
  depletion.  The solid curve shows the effect of local abundance fluctuations 
  resulting from incompletely mixed infall of D-enhanced, Fe-free material.}
  \label{fig:dvsfe}
\end{figure}

While there has not been much discussion in the literature of the factor 
$\sim 4$ range in oxygen abundances observed along lines of sight (LOS) 
within $\sim 1 - 2$~kpc of the Sun, much attention has been paid to the 
factor of $\sim 3$ range in the deuterium abundances.  The general consensus 
is that the large range in observationally-inferred deuterium abundances 
is the result of preferential depletion onto grains of D, relative to H 
\cite{depletion}.  Although preferential depletion of D is likely the correct 
explanation for most, possibly all, of the observed variation in ISM D 
abundances, there is, in fact, surprisingly little independent data in 
support of it.  For example, if depletion onto grains is the culprit, it 
might be expected that the D/H ratios would correlate (or anticorrelate) 
with the reddening or, with the amount of molecular hydrogen.  However, 
as the two panels of Figure~\ref{fig:ext} reveal, both high and low D 
abundances are found at very small values of E(B -- V) and of N(H$_{2}$) 
and, while there may be weak evidence in favor of decreasing D abundances 
at high E(B -- V), both high and low D/H values are found along the LOS 
with higher N(H$_{2}$).  This absence of a smoking gun for depletion is 
also revealed in Figure \ref{fig:dvsfe} where the D abundances are shown 
as a function of the corresponding iron abundances along the same LOS.  
While there {\bf IS} evidence of a correlation between D/H and Fe/H at 
low Fe abundances, that trend disappears for log~$y_{\rm Fe} \equiv 
10^{6}({\rm Fe/H})~\ga~0$, hinting that along those LOS with log~$y_{\rm 
Fe}~\ga~0$, deuterium may be undepleted. 

Whatever is the correct explanation for the large range in the ISM abundances 
of D and O, it is clear that the local ISM is {\bf NOT} homogeneous.  Since 
any local depletion of D onto grains has not been homogenized in the ISM, it 
is not unreasonable to suspect that there might also be some LOS along which 
infalling, D-enhanced, O-poor gas has been incompletely mixed with the processed, 
D-astrated, O-enhanced gas already present~\cite{srt}.  This latter effect is 
illustrated in Figure~\ref{fig:dvsfe} by the solid curve which shows the effect 
of unmixed, nearly primordial infall~\cite{srt} to yield higher D and lower 
Fe abundances than is predicted by the fiducial Galactic chemical evolution 
(GCE) model, shown by the star.  Abundances to the left and below the dashed 
lines may have resulted from Fe and D depletion respectively, while any D 
abundances above the dashed line may have been contaminated by infall.  These 
two competing processes, depletion onto grains and infall, complicate the 
attempt to use the data to infer the ``true'' ISM abundances of D and Fe 
(and O). 

\section{Galactic Chemical Evolution}

Unable to survive the high temperatures in stars, deuterium is destroyed 
when gas is incorporated in stars.  As gas cycles through stars and the 
stellar processed material is returned to the interstellar medium (ISM) 
of the Galaxy, the metallicity of the ISM increases while the abundance 
of deuterium decreases.  Infall to the disk of the galaxy of essentially 
unprocessed gas plays a crucial role in the chemical evolution of the 
Galaxy~\cite{infall}.  Infall of primordial (or very nearly primordial) 
gas which is metal-free (or very nearly so) and whose deuterium abundance 
is primordial (or very nearly so) dilutes the metallicity of the ISM while 
enhancing its D abundance~\cite{srt,tosi,tijana}.

\begin{figure}
  \includegraphics[height=.371\textheight]{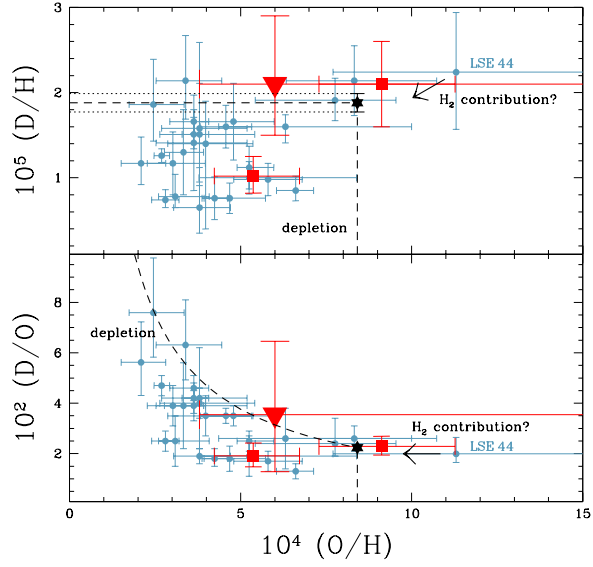}
  \caption{The upper panel shows the FUSE observed deuterium and oxygen 
  abundances, while the lower panel plots the D to O ratios as functions 
  of the oxygen abundances; from~\cite{srt}.  The stars show the ISM D 
  and D/O versus O abundances predicted by the fiducial GCE model (see 
  the text).  Abundances below and to the left of the dashed lines are 
  consistent with D and O depletion.}
  \label{fig:deplet}
\end{figure}

In Steigman, Romano, Tosi (SRT)~\cite{srt}, we considered GCE models 
with several different prescriptions for the stellar IMF and stellar 
lifetimes and predicted the ISM D astration factor and the current 
ISM oxygen (and iron) abundances.  For those GCE models consistent 
with other independent, observational data, SRT found that $1.4~\la 
f_{\rm D}~\la 1.8$ and that $7~\la y_{\rm O}~\la 12$, where $y_{\rm 
O} \equiv 10^{4}$(O/H).  For our fiducial model, SRT adopted the 
Scalo IMF~\cite{scalo} along with the Schaller {\it et al.}~\cite{schaller} 
prescription for stellar lifetimes, corresponding to $f_{\rm D} = 
1.39$ and $y_{\rm O} = 8.4$; these values are indicated by the stars 
in Figures~\ref{fig:dvsfe} and~\ref{fig:deplet}.  As noted by SRT if 
recent infall of unprocessed material were not fully mixed in the 
ISM along some LOS, there would be an anticorrelation between 
the D and O abundances along such contaminated LOS, predicting 
$y_{\rm D}^{\rm OBS} = 2.61 - 0.087y_{\rm O}^{\rm OBS}$ (the dashed
line in the upper panel of Fig.~\ref{fig:deplet}).  SRT also noted 
that since some of the dispersion observed among the D and O abundances 
may be due to systematic errors in the \hi~column densities, the D/O 
ratio would be unaffected by such errors.   For LOS contaminated by 
unmixed infall, 10(D/O)~$\equiv y_{\rm D}/y_{\rm O}$, 10(D/O)~$= 26.1 
- 0.87$ (the dashed line in the lower panel of Fig.~\ref{fig:deplet}).
 
\section{Towards An Estimate Of The D Astration factor}

\begin{figure}
  \includegraphics[height=.36875\textheight]{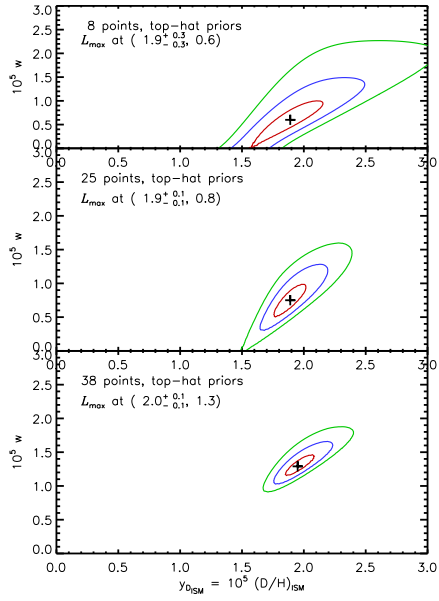}
  \caption{The 68\%, 95\% and 99\% contours in the $y_{\rm D} - w$ 
  plane ($y_{\rm D} \equiv 10^{5}$(D/H)$_{\rm ISM}$) for a top-hat 
  prior distribution for three subsets of the FUSE D/H data; see 
  the text for details.  From~\cite{pfs}.}
  \label{fig:dtophat}
\end{figure}
  
\begin{figure}
  \includegraphics[height=.3686\textheight]{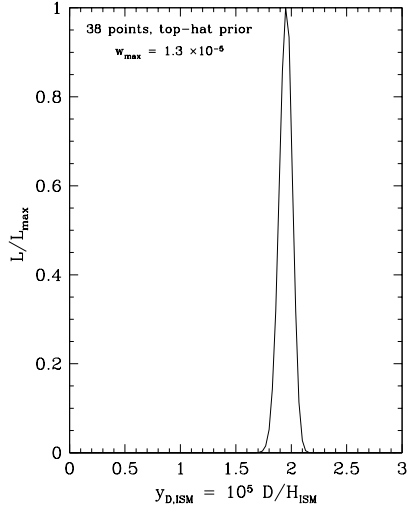}
  \caption{The likelihood distribution for the ISM D abundance 
  ($y_{\rm D,ISM} = 1.95^{+0.15}_{-0.09}$) inferred from the data 
  for 38 LOS (see Fig.~\ref{fig:dtophat}), using the top-hat prior 
  for $w$.  From~\cite{pfs}.}
  \label{fig:dlikely}
\end{figure}

Estimating the Galactic D astration factor is complicated by the possibilities
that the ``true'' ISM D abundance along some LOS may have been reduced by 
preferential D depletion onto grains and/or, enhanced by incompletely mixed 
infall of gas with the higher, primordial D abundance.  Linsky {\it et al.} 
\cite{linsky} {\bf assume} that unmixed infall makes no contribution at all so 
that, allowing for depletion, the {\bf maximum} of the ISM D/H ratios provides 
a {\bf lower limit} to the ``true'' ISM D abundance.   Linsky {\it et al.} then 
find the weighted mean of D/H for the 5 highest D abundances, concluding 
that $y_{\rm D}^{\rm ISM} \geq y_{\rm D}^{\rm MAX} = 2.17 \pm 0.017$ (or, 
including a correction for the Local Bubble D abundance, $y_{\rm D}^{\rm 
MAX} = 2.37 \pm 0.024$).  These choices correspond to astration factors 
$f_{\rm D}~\la 1.1 - 1.2$, considerably smaller than those predicted by 
the GCE models, $f_{\rm D} \approx 1.4 - 1.8$~\cite{srt}.  In contrast, SRT 
allow for the possibility that the highest observed D/H ratios may represent 
the ``true'' ISM D abundance, although unmixed infall may have contaminated 
the D abundance along some LOS and, they note that there are many more 
than 5 LOS with high D/H abundances which are equal within the errors.  
SRT find that for the 18 FUSE LOS with the highest D/H ratios ($y_{\rm D} 
\geq 1.5$), nine have $y_{\rm D} \geq 1.9$ and nine have $y_{\rm D} \leq 
1.7$, and they find for the weighted mean, $y_{\rm D}^{\rm ISM} = 1.9$.  
For the primordial abundance $y_{\rm DP} = 2.7 \pm 0.2$, this corresponds 
to $f_{\rm D} = 1.4 \pm 0.1$, in agreement with the fiducial model.

Recently, B. Fields, T. Prodanovi\'{c}, and I~\cite{pfs} have explored a different 
path to estimating $y_{\rm D}^{\rm ISM}$, using a Bayesian approach pioneered by 
Hogan, Olive, and Scully~\cite{hogan}.  In our analysis we neglect unmixed infall 
and assume that the scatter among the FUSE-observed D/H ratios is entirely due 
to preferential depletion of D onto grains, so that our estimate likely provides 
an {\bf upper bound} to the true value of $y_{\rm D}^{\rm ISM}$.  We assume 
that the gas phase D abundances (including uncertainties) are equal to the 
differences between the true ISM D abundance and corrections, $w \equiv y_{\rm 
D}^{\rm ISM} - y_{\rm D}^{\rm OBS}$, for depletion.  We adopt a prior for the 
distribution of $w$ and maximize the likelihood to find the observed abundances 
as a function of $y_{\rm D}^{\rm ISM}$ and $w$.  In Figure \ref{fig:dtophat} are 
shown the contours in the $y_{\rm D}^{\rm ISM}$ -- $w$ plane for a top-hat 
distribution for $w$ for three subsets of the data.  As the number of data 
points is increased, the contours shrink and $w > 0$ is strongly favored, 
arguing in support of depletion.  The likelihood distribution for $y_{\rm 
D}^{\rm ISM} = 1.95^{+0.15}_{-0.09}$, derived from the top-hat prior for D/H 
along 38 LOS~\cite{pfs}, is shown in Figure~\ref{fig:dlikely}.  Combining this 
estimate for (a lower bound to) $y_{\rm D}^{\rm ISM}$ with the weighted 
mean of the QSOALS D/H ratios~\cite{pettini} from Simha \& Steigman~\cite{bbn}, 
$y_{\rm DP} = 2.70^{+0.22}_{-0.20}$, results in a D astration factor $f_{\rm 
D}~\ga 1.38^{+0.13}_{-0.15}$ which, within the uncertainties, is consistent 
with all models of GCE.

\begin{theacknowledgments}
  I thank the organizers for the invitation to speak at this conference
  and the editors of these proceedings for their patience.  Much of what 
  I have presented here has resulted from collaborations with B.~D. Fields, 
  T. Prodanovi\'{c}, D. Romano, and M. Tosi, and I gladly acknowledge their 
  help and advice; I hasten to add that any errors here are mine alone.  
  My research is supported at The Ohio State University by a grant from 
  the US Department of Energy.  The work reported here was carried out 
  when I was a Visiting Professor at IAG -- USP in S$\tilde{\rm a}$o 
  Paulo, Brazil and was supported by a grant from FAPESP.
\end{theacknowledgments}


\begin{thebibliography}{99}

\bibitem{els}
R.~J. Epstein, J. Lattimer, and D.~N. Schramm, \emph{Nature},
\textbf{263}, 198 (1976); T. Prodanovi\'{c} and B.~D. Fields,
\emph{ApJ}, \textbf{597}, 48 (2003).

\bibitem{pettini}
M. Pettini et al., \emph{MNRAS}, \textbf{391}, 1499 (2008).

\bibitem{bbn}
J.~P. Kneller and G. Steigman, \emph{New J.~Phys.},
\textbf{6}, 117 (2004); G. Steigman, \emph{Int.~J.~Mod.~Phys.},
\textbf{E15}, 1 (2006); G. Steigman, \emph{Ann.~Rev.~Nucl.~Part.~Sci.},
\textbf{57}, 463 (2007); V. Simha and G. Steigman, \emph{JCAP},
\textbf{06}, 016 (2008).

\bibitem{copernicus} D.~G. York and J.~B. Rogerson \emph{ApJ}, \textbf{203} 
378 (1976); C. Laurent, A. Vidal-Madjar, and D.~G. York, \emph{ApJ}, 
\textbf{229}, 923 (1979); R. Ferlet, A. Vidal-Madjar, C. Laurent, and 
D.~G. York \emph{ApJ}, \textbf{242}, 576 (1980). 

\bibitem{linsky} J.~L. Linsky {\it et al.}, \emph{ApJ}, \textbf{647}, 
1106 (2006); C.~M. Oliveira, H.~W. Moos, P. Chayer, and J.~W. Kruck,
\emph{ApJ}, \textbf{642}, 283 (2006).

\bibitem{depletion} M. Jura, in \emph{Advances in UV Astronomy}, p.~54
[Y. Kondo, J. Mead, and R.~D. Chapman, eds., NASA, Washington] (1982); 
B.~T. Draine, in \emph{Origin and Evolution of the Elements}, p.~317 
[A. McWilliam and M. Rauch, eds., Cambridge Univ. Press, Cambridge] 
(2004).

\bibitem{srt}
G. Steigman, D. Romano, and M. Tosi, \emph{MNRAS}, \textbf{378}, 576 (2007).

\bibitem{infall}
M. Tosi, \emph{A\&A}, \textbf{197}, 33 (1988); {\it ibid} \emph{A\&A},
\textbf{197}, 47 (1988).

\bibitem{tosi}
G. Steigman and M. Tosi, \emph{ApJ}, \textbf{401}, 150 (1992); {\it ibid}
\emph{ApJ}, \textbf{453}, 173 (1995); E. Vangioni-Flam, K.~A. Olive, and
N. Prantzos, \emph{ApJ}, \textbf{427}, 618 (1994); N. Prantzos, \emph{A\&A},
\textbf{310}, 106 (1996); M. Tosi, G. Steigman, F. Matteucci, and C. Chiappini,
\emph{ApJ}, \textbf{498}, 226 (1998).

\bibitem{tijana} T. Prodanovi\'{c} and B.~D. Fields, \emph{JCAP}, \textbf{09},
003 (2008).

\bibitem{scalo} J.~M. Scalo, \emph{Fund.~Cosm.~Phys.}, \textbf{11}, 1 (1986).

\bibitem{schaller} G. Schaller, D. Schaerer, G. Meynet, and A. Maeder,
\emph{A\&AS}, \textbf{96}, 269 (1992).

\bibitem{pfs} T. Prodanovi\'{c}, B.~D. Fields, and G. Steigman, In 
preparation (2008).

\bibitem{hogan} C.~J. Hogan, K.~A. Olive, and S~.T. Scully, \emph{ApJL},
\textbf{489}, L119 (1997).

\end{thebibliography}
\end{document}